\documentclass[12pt]{article}
\pdfoutput=1
\usepackage{cite}
\usepackage{amssymb}
\usepackage{amsmath}
\usepackage{bm} 
\usepackage{graphicx}
\usepackage{color}
\usepackage{xcolor}
\usepackage{dsfont}
\usepackage{enumerate}
\usepackage{hyperref}
\usepackage{setspace}
\usepackage[latin1]{inputenc}
\definecolor{nicered}{rgb}{0.7,0.1,0.1}
\definecolor{nicegreen}{rgb}{0.1,0.5,0.1}
\hypersetup{colorlinks,citecolor= nicegreen,linkcolor= nicered}

\newcommand{\be}  {\begin{equation}}
\newcommand{\ee}  {\end{equation}}

\def\e6{E(6)}
\def\10{SO(10)}
\def\21{SA(2) $\otimes$ U(1) }
\def\321{$\mathrm{SU(3) \otimes SU(2) \otimes U(1)}$ }

\def\422{SA(4) $\otimes$ SA(2) $\otimes$ SA(2)}
\def\roughly#1{\mathrel{\raise.3ex\hbox{$#1$\kern-.75em
      \lower1ex\hbox{$\sim$}}}} \def\lsim{\roughly<}
\def\gsim{\roughly>}

\def\lsim{\raise0.3ex\hbox{$\;<$\kern-0.75em\raise-1.1ex\hbox{$\sim\;$}}}
\def\gsim{\raise0.3ex\hbox{$\;>$\kern-0.75em\raise-1.1ex\hbox{$\sim\;$}}}
\nopagebreak
\allowdisplaybreaks
\begin{document}


\newcommand{\MP}{{\sl \small Max-Planck-Institut f\"ur Kernphysik\\ \sl \small Saupfercheckweg 1, 
69117 Heidelberg, Germany}}
  \vspace*{0.5cm}
\begin{center}
 \textbf{\LARGE Toward a theory of  neutrino mass and mixing \footnote{
Based on lectures given at the Mayorana school, Modica, Sicily, July 4 - 14, 2023.
Written is a ``view of the research field" requested by the organizers.}}
 \vspace{0.8cm}\\
 Alexei Yu. Smirnov\footnote{\href{mailto:smirnov@mpi-hd.mpg.de}{\tt smirnov@mpi-hd.mpg.de}}  \\
  \MP.
\end{center}




\vspace*{0.2cm}

\begin{abstract}
Among numerous theoretical ideas, approaches, mechanisms, models  
there are probably few elements which will  eventually enter 
the true theory of neutrino masses and mixing. 
The task is to identify them. Still something conceptually important can be  missed.  
The problems of construction of the theory are outlined. 
Perspectives and possible future developments are discussed. 
\end{abstract}

\vspace*{0.8cm}



The theory of neutrino mass and mixing does not exist yet. Probably there is no sense to 
talk about such a theory separately, without connection to masses and mixing of other fermions 
as well as some other phenomena. What we call a theory of neutrino mass and mixing  
appears presently as a multi-dimensional landscape of approaches, models, 
schemes and mechanisms \cite{Feruglio:2019ybq} \cite{Ding:2023htn} 
\cite{Chauhan:2023faf} \cite{Xing:2020ijf}~\footnote{I apologize for very poor citations: space for complete 
reference list would be order of magnitude longer than the text of the paper.}. 
Possible energy scales  responsible for the neutrino mass generation spread over 50 orders of magnitude  
from sub-sub- ($\sim 10^{-20}$) eV   up to the Planck mass. 
Concerning the mixing,  the ideas range from nearly exact symmetries to anarchy.  
One, two or more extra dimensions  can be involved. 
No unique line of developments can be traced;  
various directions are still open and not much ``stuff" is excluded.  

Essentially scanning of possibilities was performed in the framework of ``QFT plus Flavor symmetries".  
Classification of the produced theoretical material can be done.
Effective field theory was used to describe the possibilities 
in terms of operators of various dimensions. 

Probably correct elements of the theory are already among numerous proposals.  
The task is then to identify them. 
And still something important can be missed. 
Below are several ``items"  which have a  chance to ``survive" 
or to play important role in the identification procedure.

\section{Beginning and the end}

The theory of mass and mixing may start and end here: 
\begin{equation}
\frac{g_{\alpha \beta}}{\Lambda} l_\alpha^T l_\beta H^T H  \rightarrow m_{\nu, \alpha \beta} = 
g_{\alpha \beta} \frac{\langle H \rangle^2}{\Lambda}, ~~~ \alpha, \beta = e, \mu, \tau ,   
\label{weinberg}
\end{equation}
where $l_\alpha$ and $H$  are the lepton and Higgs doublets \cite{Weinberg:1980bf}.   
For $g_{\alpha \beta} \sim O(1)$, the scale of new physics 
$\Lambda \sim few \times 10^{14}$ GeV,  
and Weinberg would say that's it. 
The $\beta \beta_{0\nu}-$decay should be eventually seen,  
and probably proton decay will be observed. Dark matter and energy may not be connected to 
neutrinos.  
That's it, unless we will 
discover some new physics at lower energy scales, new particles,     
lepton number violating processes, detect gravitational waves from the phase 
transitions associated to the violation of lepton number \cite{Brdar:2018num}, {\it etc.}  
Theory of $g_{\alpha \beta}$ couplings can be elaborated 
 without real tests.\\ 

The opposite in many aspects scenario is $\nu$MSM,  which is SM $+ 3 \nu_R$ plus idea of minimality.  
The $\nu$MSM is essentially the phenomenological model with 
UV completion at the string-Planck 
scale such that new physics below $M_{Pl}$ does not exist  
\cite{Asaka:2005pn}, \cite{Asaka:2005an}, \cite{Boyarsky:2009ix}.
The  model  pretends to explain  all the observations: Smallness of neutrino mass 
is obtained by combination of the low (EW) scale  seesaw and smallness of the Dirac Yukawa couplings.     
The lepton asymmetry of the Universe is generated  via oscillations of the RH neutrinos 
 $\nu_{2R}$ and  $\nu_{3R}$ and then converted by sphalerons to the baryon  asymmetry. 
The RH neutrinos $\nu_{2R}$ and  $\nu_{3R}$  may be produced in  $B-$decays (Br $\sim 10^{-10}$) 
and  tested at SHiP.  Higgs inflation was invented. 
$\nu_{1R}$ being the DM particle  is produced  via the resonance conversion 
of active neutrinos in the Early Universe, it  
decouples from generation of 
light neutrino mass. The mass spectrum  can be supported by symmetry  
with starting point of the  degenerate pair  $\nu_{2R}$, $\nu_{3R}$ and massless state $\nu_{1R}$. 
The model is  still alive and can  be tested. A possibility of Grand Unification is problematic.  \\

The right handed (RH) neutrinos, $\nu_R$,  
should be present in the SM of particle physics.  
Why not,  if other SM fermions do have RH components?
Their existence is justified in plausible extensions of SM:
gauged $L - R$, Pati-Salam symmetry,  $SO(10)$. 
The interaction of $\nu_R$ with $\nu_L$ and Higgs 
(with coupling $g$) is not forbidden,  and 
consequently,  the  Dirac mass term should be  generated: 
\begin{equation}
y \bar{l} \nu_R H  + h.c.  \rightarrow  y \bar{\nu}_L \nu_R \langle H \rangle + h.c. 
\rightarrow 
m_D =  y \langle H \rangle.
\label{dirac}
\end{equation}
If the observed neutrino mass $m_\nu = m_D$,  {\it i.e.} neutrinos are the Dirac particles, 
the required coupling 
$y \simeq 2 \cdot  10^{-13}$ is uncomfortably (for us but may not for Nature) small. 

The term (\ref{dirac}) could  be forbidden if $\nu_R$ 
has some quantum number which distinguishes  
it from other SM particles. In this case more appropriate  to 
call $\nu_R$  new neutral lepton not associated to fermionic generations rather than 
the  RH neutrino.

\section{Right handed neutrinos are the key}

Existence and properties (interactions) of  $\nu_R$ are 
the key  points of theory of $\nu$ mass and mixing. 
Various possibilities are related to their nature and  scale of masses  $M_R$.
If $\nu_R$ is a Majorana particle and has large Majorana mass
$M_R \gg m_D$,  the seesaw type I mechanism is realized 
\cite{Minkowski:1977sc} \cite{Gell-Mann:1979vob} \cite{Yanagida:1980xy}
\cite{Mohapatra:1979ia}, which is  behind the operator (\ref{weinberg}) with 
$\Lambda  =  M_R \simeq (10^9 - 10^{14})$ GeV. 
If $M_R \simeq M_{Pl}$, the neutrino mass generated 
via seesaw, $m_\nu \sim 10^{-5}$ eV \cite{Akhmedov:1992hh},  
is  much smaller than the observed one.  
Some other mechanism should give the main contribution,
e.g. the radiative, seesaw type II, environmental  
mechanisms. Here the seesaw acts as the mechanism of suppression of 
the Dirac mass term effect. Among radiative mechanisms the scotogenic one 
\cite{Tao:1996vb}, \cite{Ma:2006km} with 
new neutral lepton $S$ and  new scalar doublet 
odd with respect to $Z_2$ symmetry 
looks attractive establishing connection to the Dark Matter in the Universe.

The RH neutrinos may have no Majorana mass term but couple to new leptons $S$ 
with Majorana mass $M_S$:  
\begin{equation}
M_D \bar{\nu}_R S + \frac{1}{2} M_S SS  + h.c.
\label{eq:sterile}
\end{equation} 
If $M_S \ll M_D$, then $\nu_R$ and $S$ form pseudo-Dirac neutrino
with mass $\approx M_D$.  
By adding the term (\ref{dirac}) to (\ref{eq:sterile}), the inverse seesaw can be realized for 
usual neutrinos \cite{Mohapatra:1986bd}. 
Here the $\nu_R - S$  system plays the role of the right handed neutrino with 
an effective mass $\Lambda = M_R^{eff} \approx - M_D^2/M_S$.
If $M_S$ is very small  (e.g., in the keV range) large value  of $M_R^{eff}$
can be obtained for $M_D$ at the LHC scale.
Now $\Lambda$ is not a fundamental but a fictitious scale composed of two much smaller scales.

For $M_S \gg M_D$ the Majorana mass of $\nu_R$ 
is generated by higher scale seesaw 
$M_R = - M_D^2/M_S$. In this way, for active neutrinos
the double seesaw is realized. A possibility 
 $M_D \simeq M_{GUT}$, $M_S \simeq M_{Pl}$, which gives
$M_R \simeq 10^{14}$ GeV required by the usual seesaw, looks very suggestive. 

The Majorana mass of $\nu_R$ may have non-trivial dynamical origin, for instance,  
$M_R = h_\Delta \langle \Delta_R \rangle$, where $\Delta_R$ 
is the   $SU(2)_R$ Higgs triplet  in the L-R symmetric models or
$M_R = h_\phi \langle \sigma \rangle$,  where  $\sigma$ is the gauge
singlet. It can originate from condensate of new strongly interacting sector.

Singlets $S$,  which couple to  $\nu_R$, can  
interact with new scalar and vector bosons, thus forming  
whole new sector - the Dark sector of Nature. 
Then $\nu_R$ plays the role of ``portal" to this sector. 

Not only masses but also mixing can be related to properties of $\nu_R$. 
Many singlets $S$  
organized in special way may exist which can produce large mixing or kind 
of random  mixing patterns of light neutrinos.

Smallness of coupling $y$ in  (\ref{dirac})   neutrino mass can be related to localization of 
$\nu_R$ is extra dimensions which differs from localization
of particles with non-zero EW charges. As a result, the overlap of WF 
of $\nu_R$ and $\nu_L$, and consequently, masses are strongly suppressed. E.g. $\nu_R$ can propagate 
in whole extraD space while $\nu_L$ is localized in 3D brane 
\cite{Arkani-Hamed:1998wuz}. 
Or $\nu_L$ and $\nu_R$ are localized  on different branes and  
overlap of their wave functions is exponentially suppressed \cite{Grossman:1999ra}.

The RH neutrino masses may be the origins of  the EW scale
(``Neutrino option") \cite{Brivio:2017dfq}:  
Both the Higgs mass term and quartic coupling  (absent at tree level)
are generated by neutrino ($\nu_L - \nu_R$) loops. This requires  
$M_R = (10^7 - 10^9)$  GeV, and $y =  (10^{-6} - 10^{-4.5})$. 

\section{Environmental mass: VEV versus EV}

The neutrino mass may have an environmental  origins 
being related to the Dark matter or/and Dark energy \cite{Fardon:2003eh} in the Universe,  as well as   
to new physics at very low energy scales.  Indeed, the oscillation results can be explained by
any  term in the Hamiltonian  of evolution with  $1/E$ dependence.    
This can be potential produced by particles of background if these 
particles and mediators of interactions are sufficiently light. 

In the standard model neutrino mass is of vacuum origins generated by neutrino interaction 
with Higgs field in its lowest energy state - VEV. If the vacuum mass is suppressed by Planck scale seesaw,  
the dominant contribution may come from the neutrino interactions (coupling $g$)
in the background composed of scalars particles $\phi$. At high number density, $n_\phi$, 
the background can be treated as classical field  
- the expectation value of  the field operator in the coherent state of scalar background: $\langle \phi \rangle 
\simeq \sqrt{n_\phi/m_\phi}$, where $m_\phi$ is the mass of scalar \cite{Berlin:2016woy}. 
Then the effective  neutrino mass 
equals $m_\nu^b = g \langle \phi \rangle \propto  g \sqrt{n_\phi/m_\phi}$. 
Oscillations of ultra-relativistic neutrinos 
are determined  by masses squared:  
$(m_\nu^b)^2/2E = g^2 n_\phi/(2 E m_\phi) $. The same result can be obtained considering   
refraction:  neutrino elastic forward scattering on 
scalar bosons $\phi$ \cite{Choi:2019zxy} \cite{Choi:2020ydp} \cite{Sen:2023uga}.  

Locally, the difference between the vacuum and refraction masses 
(for scalar DM) is not so significant: 
In one case the mass is due to VEV,  in another one -- due to EV (Expectation Value). 
In contrast to the vacuum mass the refraction mass being proportional to 
$n_\phi$ depends on space-time coordinates, as well as on energy. 
Additional time dependence,  $\cos m_\phi t$, appears for non-relativistic coherent state of $\phi$. 
In model \cite{Sen:2023uga} the values of  parameters are $m_\phi  \lesssim 10^{-10}$ eV, $g \lesssim 10^{-10}$ and 
$m_\chi  \lesssim 10^{-4}$ eV (mass of mediator). 

It is not completely clear if refraction mass can 
explain oscillation results without contradicting other,  in particular, cosmological observations.  
It does not provide (add) any new insight into mixing and mass spectrum. 
But it is important to search for the experimental   
consequences of refraction mass:  
Discovery of space - time dependence 
of the oscillation parameters will shed light not only on nature of 
neutrino mass but also Dark Matter.




Yet another interesting possibility exists: neutrino condensate 
$\Phi_{\alpha \beta}  =  \langle \nu_\alpha^T \nu_\beta \rangle$  can be formed  
due to non-perturbative gravitational interactions 
(gravitational $\theta$-term) 
in analogy with the quark condensate \cite{Dvali:2016uhn}. 
Neutrino mass generated by this condensate equals  
$m_{\alpha \beta} \simeq \Phi_{\alpha \beta}$.

\section{Mass, mixing and symmetries}

Ideas about  the mass-mixing connection range 
from strict relationships to complete decoupling. 
Neutrinos have  the weakest mass hierarchy (if any) among fermions.  
For normal ordering one finds  $m_2/m_3 \geq \sqrt{\Delta m_{21}^2/ \Delta m_{31}^2} \simeq 0.17$. 
Large lepton mixing can be due to this weak hierarchy, 
via the  Gatto-Sartori-Tonin type relation: $\theta \simeq \sqrt{ m_2/m_3}$ 
\cite{Gatto:1968ss}.  
It  can be realized even better if neutrino spectrum is non-hierarchical. 
As a general guideline,  mixing is related to  different mass hierarchies 
of the  upper and down fermions, while difference of quark and lepton mixings 
is related to smallness of neutrino mass.  

Decoupling of mass from mixing looks very counterintuitive and non-trivial. 
The approximate decoupling can be achieved since   
certain relations (e.g. equalities) between elements 
of the mass matrix give mixing independently on the size of elements.  
This can be a  consequence of certain symmetry.    
In turn, the ratios of mass terms  from different relations 
fix ratios of  mass eigenvalues. \\

\noindent
In the first approximation the lepton mixing is described by the TBM matrix \cite{Harrison:2002er}  
with  deviations of the order of Cabibbo angle - ``Cabibbo haze" \cite{Datta:2005ci}. 
This can be accidental and even consistent with anarchy 
\cite{Hall:1999sn}.  In turn, the ``haze" can be random or organized as 
in the quark - lepton complementarity approach. 
Alternative point of view is that the TBM pattern is non-accidental and  
originates from  certain broken symmetry.  
Residual symmetry approach  was elaborated to realize the latter. 
 
The approach is based on (i) decoupling of masses and mixing, and (ii) 
different intrinsic symmetries of the $\nu-$ and $l-$ mass matrices. 
Indeed, TBM can not be related to mass ratios, and therefore implies decoupling of masses and mixing. 
The mixing appears as a result of different ways of the flavor symmetry breaking by flavon fields 
in the  neutrino and charged lepton  (Yukawa) sectors. For this ``sequestering" of the 
corresponding flavons required.  In turn, this difference can be related  
to nature of neutrino and charge leptons masses: Majorana versus Dirac, and  
to different flavor symmetry charges (representations) of the
RH neutrinos and charged leptons. 
The flavor symmetry is broken down to residual symmetries (different for $\nu$ and
$l$) which can coincide with intrinsic symmetries of the corresponding
mass matrices. 
One can proceed in the opposite way and reconstruct 
the flavor symmetry(ies) from the intrinsic symmetries and 
the TBM mixing. The fact, that rather simple symmetries like 
$S_4$ are obtained \cite{Lam:2008rs}, indicates that something substantial can be 
in this approach.

Realization of the residual symmetries program in specific gauge models 
turns out to be complicated  and not convincing with {\it ad hoc} introduced structures, 
large number of parameters, {\it etc}. 
The dilemma is  ``wrong prediction versus no predictions''. 
The flavor charge assignment (set of free discrete parameters) 
has no clear logic, some low dimensional   
representation are absent 
(the missing representations problem).
Mass hierarchy arises from a kind of Froggatt-Nielsen mechanism \cite{Froggatt:1978nt}:  
Hierarchy of mass terms is due to high dimensional non-renormalizable  
operators with products of different numbers of  flavon fields.  
Discrete symmetries provide restricted possibilities to explain also  masses  
and usually lead to degenerate or partially degenerate  spectra.

Generic problem is  that masses are functions of Yukawa couplings 
and VEV's of flavon fields $\langle \phi_\alpha \rangle$:   
\begin{equation}
m = F(Y,  \langle \phi_\alpha \rangle),  
\end{equation} 
which follow from independent sectors of theory: 
$Y$ - from the Yukawa sector, while $\langle \phi_\alpha \rangle$ from  
the scalar potential. 
Here $F$ is determined by mechanism of mass generation. 
Potentially, supersymmetry can establish 
connections since both sectors originate from the superpotential. 
To get TBM, parameters of  these two sectors  should be  correlated, 
which requires  auxiliary symmetries, additional fields, {\it etc}.\\ 

\noindent
Modular symmetry as the  flavor symmetry 
was expected to resolve the problem of the ``traditional" symmetries discussed above 
\cite{Feruglio:2019ybq}, \cite{Ding:2023htn},
\cite{Trautner:2023xai}. 
It is motivated by  string theory, and 
therefore unavoidable if we believe in 
strings. The symmetry is related to compactification of extra dimensions.  
Primary it is realized on the moduli fields $\tau$ which describe geometry of the compactified space. 
New elements in the model building are 
(i) transformations:  appearance  of the weight factor in transformations;   
(ii) Yukawa couplings: the couplings are  modular forms -   
non-linear functions of moduli fields which compose multiplets and 
transform under representation of finite symmetry group $\Gamma_N$.  
Invariance condition for weights  
gives additional restrictions, forbids some mass terms, leads to  texture zeros.

Models based on modular symmetries are not motivated by TBM, so the approximate TBM appears here accidental. 
The goal was to reduce number of parameters, make theory  
more predictive, and also connect masses and mixing. 
For fixed level $N$, which fixes the discrete symmetry group,  
the fit parameters are the  VEV of moduli fields (continuous complex number), 
weights of matter multiplets and Yukawas, the overall couplings at the 
invariant interaction terms. 
The models allow to reproduce the observed mixing angles and mass splits and 
to predict the absolute values of masses, and  CP-phases. Typically  weak hierarchy 
and  often quasi-degenerate mass spectrum are predicted. 
For some specific points of moduli space the hierarchical 
mass spectrum can appear, and  it seems,  
$\tau \simeq i$ plays special role. 
%
Here model building is reduced to symmetry building to match the data. \\

\noindent
There is certain similarity of the residual (``traditional") 
symmetry and modular symmetry approaches.  
Indeed, the considered finite modular symmetry subgroups  are isomorphic to the groups 
$A_4$, $S_4$, $A_5$ ... (determined by the level $N$)  
used in the residual symmetry approach.  Yukawas are  modular forms 
$Y(\tau) = [Y_1(\tau),~ Y_2(\tau),~ Y_3(\tau)  ...   ]$,  while in usual approach  the effective Yukawa 
couplings depend on VEV's,  
$\langle \phi \rangle = [\langle \phi_1 \rangle, \langle \phi_2 \rangle, \langle \phi_3 \rangle ...]$, 
or products of VEV's of flavon  fields  $Y^{eff} = y (\Pi_i \langle \phi_i \rangle /\Lambda^n)$. 
But one can establish the correspondence: 
\begin{equation}
Y(\tau) \leftrightarrow  \frac{1}{\Lambda}\langle \phi \rangle .
\label{eq:eqcompar}
\end{equation}
Yukawa couplings with different weights can be used, which  
can be constructed as products of modular forms of lower weight: e.g.,  
$Y^{2n} \sim (Y^{n})^2$. Then the correspondence is 
$$
Y^{2n} \leftrightarrow \frac{1}{\Lambda^2} (\langle \phi \rangle)^2, ~~ etc. 
$$
For $\langle \phi \rangle  \ll \Lambda$ the Froggatt-Nielsen  mechanism 
is realized which means that higher  weight 
modular forms can produce hierarchy of masses.  
If 
$$\langle \phi_1 \rangle : \langle \phi_2 \rangle : \langle \phi_3 \rangle   ...  
= Y_1(\tau) : Y_2(\tau) : Y_3(\tau)... , 
$$ 
the ``traditional" flavon approach can reproduce results of the modular symmetry approach.   
The  advantage of modular symmetries is that     
components of $Y-$mutiplets,  $Y_i(\tau)$,  are fixed by group parameters: level $N$, and weight $k$,  
as well as $\tau$. 
In contrast, flavon VEV's depend on parameters of potential 
and (in most of the cases) are not controlled by the flavor symmetry. 

In the minimal model only one moduli and therefore only one continuous 
complex number (moduli VEV) is involved; the rest is determined 
by structure of symmetry: level, representations, weights and still arbitrary constants 
in front of different terms. However, minimal versions of models do not work well. Additional freedom should  
be introduced by using two or more moduli, flavons,  etc. 

Furthermore, complete and consistent top-down construction based 
on heterotic string theory compactified on orbifold 
leads to the ``eclectic"  flavor symmetries \cite{Trautner:2023xai}. They include 
simultaneously and in non-trivially  unified way
symmetries $G_{traditional}$ and  $G_{modular}$ as well as
discrete $R-$symmetry, $G_R$,  and CP-symmetry.
The K\"{a}hler potential  (kinetic terms) introduces additional parameters and  freedom.  
Applications of the modular symmetries is still in 
explorative phase.




\section{Mixing from the Darkness}

Theory of neutrino mass  and mixing should be  constructed together 
with theory quark masses and mixing. 
GUT in some version should exist:   
nothing better than GUT was proposed for BSM physics. 
No theory of quark masses and mixing exists in spite of the fact that information 
in the quark sector is complete. What one can expect for leptons? 
More modest task is to understand the difference of quark and lepton  mixings. 
The mixing patterns of quarks and leptons are strongly 
different but still can be related. The 1-2 and 2-3 quark and lepton mixing 
angles sum up approximately to the maximal mixing angle $\pi/4$. 
This Quark-Lepton complementarity (QLC) \cite{Minakata:2004xt} can be formalized by the product 
\begin{equation}
U_{PMNS}  \approx V_{D}^\dagger U_X, 
\label{eq:compliment}
\end{equation}
where $V_D \simeq V_{CKM}$ and 
$U_X \simeq U_{TBM}, U_{BM} $. Eq. (\ref{eq:compliment}) implies that $V_D$  and $V_{CKM}$ 
emerge from a common sector of  ``the CKM physics",  again implying 
q-l symmetry, or  unification and  GUT.
In turn, $U_X$ follows from the Dark sector coupled to usual neutrinos via 
the $\nu_R$ portal. 
The relation (\ref{eq:compliment})  led to prediction
$\sin \theta_{13} \approx \sqrt{1/2} \sin \theta_C$ 
\cite{Giunti:2002ye} \cite{Minakata:2004xt} 
as well as to prediction for the Dirac CP phase provided 
that $V_D$ is the only source of CP violation.
The dark sector at the String - Planck scale is responsible 
for large lepton mixing and smallness of neutrino mass 
via the double seesaw. 

This framework opens another possibility to realize flavor symmetries. 
It is much easier to introduce these symmetries in the Dark (SM singlet)  sector, $S$,  and transfer 
the information about mixing to the visible sector via  common  
basis fixing symmetry, e.g. $Z_2 \times Z_2$ \cite{Ludl:2015tha}.  
A possible setup is $SO(10)$ with $\nu_R$  portal to Dark sector 
\cite{Smirnov:2018luj} (see Fig. 1).  
The flavor symmetry is broken spontaneously or explicitly 
down to basis fixing symmetry in the portal and visible sectors. 
Symmetry breaking effects are  small being 
suppressed by ratio of the GUT and Planck scales at least.  
Still theory of CKM physics should be developed. 

What is the nature of Dark sector: is this the    
``parallel world",  
or it is completely asymmetric to the visible sector? 
Is this sector above GUT energy scale and  
should be unified with GUT before unification with gravity? 
In fact, string theory supports existence of the Dark sector.  
\begin{figure}[!t]
\begin{center}
\includegraphics[width=0.7\textwidth]{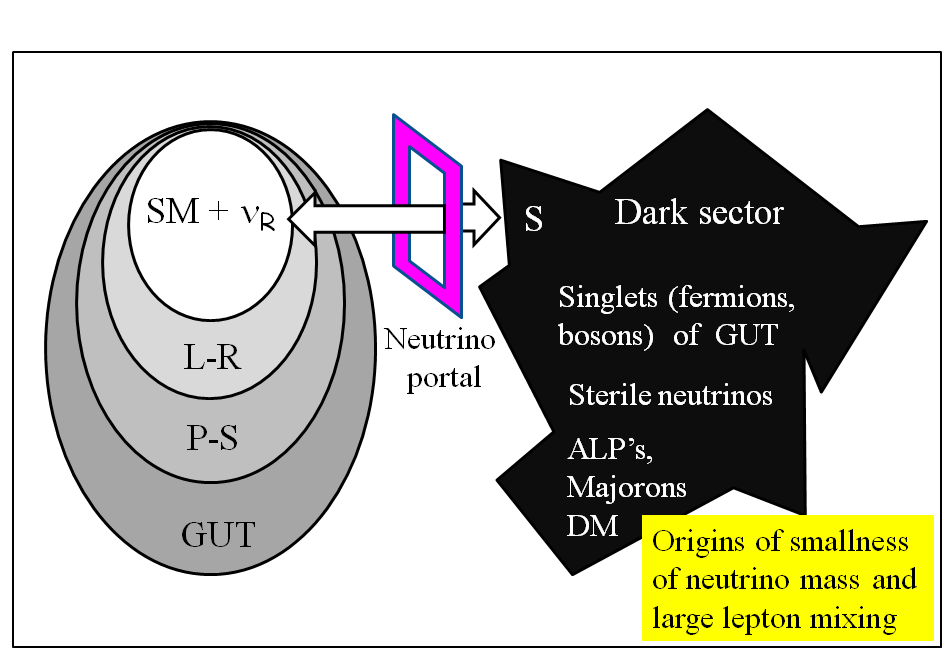}
\end{center}
\caption{
Possible set-up for the neutrino mixing from the Dark sector.}
\label{fig:Delmeff_g_mphi}
\end{figure}
Similar construction can be realized for low scale (10 - 100 TeV) 
Dark sector using inverse seesaw,  
e.g.,  in the framework of the L-R symmetric models.


\section{Sterile pollution}

The eV-scale sterile neutrinos $\nu_S$ mixed with  active neutrinos may exist, 
as indicated by some experiments  \cite{Acero:2022wqg}. For $m_4  = (1 - 3)$ eV and   
$\sin^2 2\theta_{eS} \approx 4(m_{eS}/m_S)^2 \simeq  0.02 - 0.1$, 
the corrections to the mass matrix of active neutrinos after decoupling of sterile neutrino, 
$$
\delta m_{ee} \approx \sin^2\theta_{eS} m_4 \sim (0.005 - 0.07) \, \, {\rm eV},  
$$
are  at the level of the largest elements ($\approx$ 0.03 eV) of  the $3\nu$- mass matrix.  
That is,  the  effect of $\nu_S$  is not a small perturbation of 
the $3\nu$-picture and possible flavor symmetries.
Oscillation data require cancellations and  
fine tuning in the full mass matrix, and  therefore $\nu_S$ 
should be included into theory and symmetry 
constructions from the beginning. 
At the same time, mixing with $\nu_S$ allows to enhance lepton mixing, to explain the difference of the quark 
and lepton mixings, and  even generate the TBM mixing,  if 
$m_{e S}$, $m_{\mu S}$, $m_{\tau S}$  have certain relations (symmetry) 
\cite{Balaji:2001ns}. 


\section{Problems and perspectives}

Where are the origins of the problems of building the theory? 
Are we mislead by the data:  
is our interpretation of the oscillation data correct? 
First step is to  play with observations and to search for regularities. 
There are many observables and some qualitative regularities. 
Among those  are the  $b - \tau$ unification,  
unification of couplings of the third generation,  {\it etc.}
But there is no  exact and simple relations between observables 
which would allow to reduce substantially number of free parameters   
and give certain hints of the underlying theory. 
There is nothing  like the Balmer series in atomic physics. 
Maybe  the situation is closer to nuclear physics and nuclear spectra. 
Probably one should use artificial intelligence  
to search for regularities or conclude about their absence.

It seems, ``One step - single framework"  constructions do not work. 
Two or more different and independent contributions to mass matrices can be present. 
And even two contributions with exact symmetries 
can produce random effect in the sum, especially if the number of degrees of freedom is small.  
This can be tested by certain mathematical tools. 
Recall that the proton mass gets contribution from QCD, electromagnetic interactions and quark masses. 
Smallness of neutrino mass means that independent 
contributions from all energy scales up to the Planck scale become relevant. 
Important task is then to ``clean up the data from pollution" 
of non-leading effects. 
As it was marked, mixing with sterile neutrinos can modify 
substantially the mass matrix, and consequently,   the mixing of active neutrinos. 
Probably  we should be satisfied with some qualitative relations and regularities,  
and ignore the ``haze" of unexplained corrections. \\  

What are  perspectives in the field? 
Which progress 
the results of forthcoming experiments can provide?  
That includes establishing mass hierarchy, 
determination of CP phase, high precision measurements of known oscillation parameters
(with reservation that our theoretical predictions have not achieved high accuracy) \cite{Huber:2022lpm}. 
In fact, we already explored in advance possible impact 
of different outcomes of future experiments 
on theory. In particular,  the  analysis of present situation
was performed in two modes (of possible mass ordering): 
normal and inverted. Some models
may be excluded and  parameter space - restricted. 
Determination of mass ordering will certainly help. 
Ideas behind the ordering range  from  
fundamental principles  and symmetries,  to  accidental: selection of values of parameters.
Establishing NH will testify for the see-saw, 
quark-lepton similarity or symmetry, unification.
Inverted ordering means strong degeneracy 
of two heavy states, and consequently, a symmetry. 
It indicates the structure Pseudo-Dirac pair plus very light  Majorana (or Weyl) neutrino. 
The $\nu_1$ and $\nu_2$ states form quasi-degenerate structure imposed by flavor symmetries, 
e.g.  broken $L_e - L_\mu - L_\tau$. 
Some special values of the CP phase  like $0, \pi, \pm \pi/2$
are very suggestive. 

Knowledge of the overall scale of neutrino masses, 
$m_\nu  \simeq  \sqrt{\Delta m_{31}^2}  = 0.045~ {\rm eV}$,   
does  not give any immediate insight 
into theory being related to some free parameters of specific models. 
But it is crucial for determination of type of mass spectrum: hierarchical,
non-hierarchical, quasi-degenerate, and this is, indeed, very important for theory. 
Quasi-degenerate spectrum is already disfavored 
and further progress is expected. 

There is strong dependence of structure of mass matrix on the unknown 
Majorana phases. In turn,  different patterns of mixing require different 
underlying symmetries. But it will be very difficult to determine the phases.  

What else should be measured and searched  to achieve the progress? 
%
Certainly we need  to 
\begin{itemize}

\item  
further strengthen the bounds on  mixing with  steriles,  thus removing possible ``pollution"; 

\item
tests of nature of neutrino mass: searches for energy 
and space-time dependencies of the oscillation parameters.   

\end{itemize}

As far as theory developments are  concerned, we should proceed further with  exploration of 
flavor symmetries (other realizations, 
modifications of symmetry, other applications).  
Studies of modular symmetries in whole string framework may reveal 
for simple versions the required corrections e.g.  from K\"{a}hler potential. 
Or maybe modular-like symmetries without strings can be elaborated. 
If this does not work some attempts can be made to 
go beyond common framework based on  QFT - flavor 
symmetries - symmetry breaking. May be more can be learned from  Swampland? 
Anyway the line of research GUT - Planck looks very appealing.

Formaly,  mathematical  structures  we use  
do not map onto the data  in a convincing way. With large number of parameters 
almost any formalism can be used to describe the data.  
Large number of parameters  in the first step of developments
should not discourage, provided that something new -    
new particles, interactions, dynamics are predicted and can be  tested. 

Model building should be computerized. Once principles and framework are determined
the programs should be developed to produce viable (consistent with observations) 
models instead of writing hundreds  of separate  papers on models which differ by  level $N$, 
weight prescriptions for the matter fields and Yukawas, number of moduli fields.

As an alternative, low and very low scale physics responsible for neutrino masses 
looks interesting.  Neutrinos are in between the two ``deserts": the high energy 
and low energy ones,  and can provide the key  progress in probes of both.

Connection with other phenomena can be  decisive.   
Any further discovery will affect the field and some findings may lead 
to breakthrough in our understanding. Actually, almost any BSM phenomenon:
new particles, sterile neutrinos, Dark mater, proton decay, {\it etc.}, 
may  shed some light on origins of neutrino mass and mixing. 


\section*{Acknowledgments}

The author acknowledge A. Trautner for useful discussions.

\end{document}